\documentclass[pre,twocolumn,superscriptaddress]{revtex4-1}
\usepackage{graphicx}
\usepackage{amsmath,amssymb}
\usepackage{dcolumn}
\usepackage{url}
\usepackage[colorlinks=true,allcolors = blue]{hyperref}

\newcommand{\press}{{\mathfrak p}}
\newcommand{\energy}{{\mathfrak e}}
\newcommand{\prob}{{\mathcal P}}

\newcommand{\leibnizd}{\mathrm{d}}
\newcommand{\dr}{\leibnizd r}
\newcommand{\dtheta}{\leibnizd\theta}
\newcommand{\dtr}{\leibnizd^3\rvec}
\newcommand{\dmu}{\leibnizd\mu}
\newcommand{\dgamma}{\leibnizd\gamma}

\newcommand{\dt}{\delta t} % for time step

\newcommand{\myvec}[1]{{\bf #1}}
\newcommand{\pvec}{\myvec{p}}
\newcommand{\Pvec}{\myvec{P}}
\newcommand{\Pbox}{\myvec{P}_{\text{box}}}
\newcommand{\kvec}{\myvec{k}}
\newcommand{\drvec}{\Delta\myvec{r}}
\newcommand{\DRvec}{\Delta\myvec{R}}
\newcommand{\rvec}{\myvec{r}}
\newcommand{\Evec}{\myvec{E}}

\newcommand{\Dr}{\Delta r}

\newcommand{\betastar}{\beta^\star}
\newcommand{\rstar}{r^\star}

\newcommand{\rcut}{r_{\text{cut}}}
\newcommand{\kcut}{k_{\text{cut}}}

\newcommand{\epssub}[1]{\epsilon_{#1}}
\newcommand{\epsr}{\epssub r}
\newcommand{\epsb}{\epssub b}
\newcommand{\gK}{g_{\text{K}}}
\newcommand{\gC}{g_{\text{C}}}
\newcommand{\phipol}{\phi_{\text{pol}}}

\newcommand{\Nmap}{{\cal N}_m} % water bead mapping number
\newcommand{\Vm}{V_m} % molar volume of water
\newcommand{\NA}{N_A} % Avogadro's number

\newcommand{\gref}{g^{\text{ref}}}
\newcommand{\phibond}{\phi^{\text{int}}}

\newcommand{\rhom}{\rho_m}
\newcommand{\Nm}{N_m}
\newcommand{\kd}{k_d}

\newcommand{\kB}{k_\text{B}}
\newcommand{\kT}{\kB T}
\newcommand{\lB}{l_{\text{B}}}
\newcommand{\rc}{r_c}
\newcommand{\lD}{\lambda_{\text{D}}}

\newcommand{\Gwater}{\Gamma_{\text{water}}}
\newcommand{\Gions}{\Gamma_{\text{ions}}}

\newcommand{\myav}[1]{\langle{#1}\rangle}
\newcommand{\myrav}[1]{\langle{#1}\rangle_r}
\newcommand{\myavC}[1]{\langle{#1}\rangle}
\newcommand{\myavK}[1]{\langle{#1}\rangle}

\newcommand{\Fpolar}{F_{\text{polar}}}
\newcommand{\Fnonpolar}{F_{\text{non-polar}}}

\newcommand{\rhobar}{{\overline\rho}}
\newcommand{\rhoitot}{\rhobar_i}
\newcommand{\rhoio}{\rhoitot^{(1)}}
\newcommand{\rhoiw}{\rhoitot^{(2)}}
\newcommand{\rhobarions}{\rhobar_{\text{ions}}}

\DeclareMathOperator{\erf}{erf}
\DeclareMathOperator{\erfc}{erfc}

\newcommand{\half}{\frac{1}{2}}
\newcommand{\smallhalf}{{\textstyle\half}}

\newcommand{\PthreeM}{P$^3$M}

\newcommand{\nm}{\text{nm}}

\newcommand{\code}[1]{{\sc #1}}
\newcommand{\DLMESO}{\code{dl\_meso}}

\newcommand{\role}{r\^{o}le}

\newcommand{\latin}[1]{{\itshape #1}}
\newcommand{\eg}{\latin{e.\,g.}}
\newcommand{\ie}{\latin{i.\,e.}}
\newcommand{\etal}{\latin{et al.}}
\newcommand{\via}{\latin{via}}
\newcommand{\cf}{\latin{cf.}}
\newcommand{\mutmut}{\latin{mutatis mutandis}}

\newcommand{\Eq}[1]{Eq.~\eqref{#1}}
\newcommand{\Eqs}[1]{Eqs.~\eqref{#1}}
\newcommand{\Fig}[1]{Fig.~\ref{#1}}
\newcommand{\Refcite}[1]{Ref.~\cite{#1}}
\newcommand{\Refscite}[1]{Refs.~\cite{#1}}
\newcommand{\Table}[1]{Table~\ref{#1}}
\newcommand{\Sec}[1]{Sec.~\ref{#1}}
\newcommand{\Appendix}[1]{Appendix~\ref{#1}}

\newcommand{\partFig}[2]{Fig.~\hyperref[#1]{\ref*{#1}#2}} % version with nicer hypertext link

\newcommand{\DS}{DS}%or DSM?
\newcommand{\WinoDim}{WinO-DIM}
\newcommand{\WinODS}{WinO-DS}
\newcommand{\DSq}[1]{DS-q#1} % or {DSq#1}? or {DS#1}?

\begin{document}

\title{Polarisable soft solvent models with applications in dissipative particle dynamics }

\author{Silvia Chiacchiera}
\email{silvia.chiacchiera@stfc.ac.uk}
\affiliation{Scientific Computing Department, UKRI Science and Technology Facilities Council, Daresbury Laboratory, Sci-Tech Daresbury, Warrington WA4 4AD, UK.}

\author{Patrick B. Warren}
\email{patrick.warren@stfc.ac.uk}
\affiliation{The Hartree Centre, UKRI Science and Technology Facilities Council, Daresbury Laboratory, Sci-Tech Daresbury, Warrington WA4 4AD, UK.}

\author{Andrew J. Masters}
\email{andrew.masters@manchester.ac.uk}
\affiliation{Department of Chemical Engineering, University of Manchester, Manchester M13 9PL, UK.}

\author{Michael A. Seaton}
\email{michael.seaton@stfc.ac.uk}
\affiliation{Scientific Computing Department, UKRI Science and Technology Facilities Council, Daresbury Laboratory, Sci-Tech Daresbury, Warrington WA4 4AD, UK.}

\date{April 12, 2024} % \date{\today}

\begin{abstract}
  We critically examine a broad class of explicitly polarisable soft solvent models aimed at applications in dissipative particle dynamics.  We obtain the dielectric permittivity using the fluctuating box dipole method in linear response theory, and verify the models in relation to several test cases including demonstrating ion desorption from an oil-water interface due to image charge effects.  We additionally compute the Kirkwood factor and find it uniformly lies in the range $\gK\approx0.7$--0.8, indicating that dipole-dipole correlations are not negligible in these models.  This is supported by measurements of dipole-dipole correlation functions.  As a consequence, Onsager theory over-predicts the dielectric permittivity by approximately 20--30\%.  On the other hand, the mean square molecular dipole moment can be accurately estimated with a first-order Wertheim perturbation theory.\\[3pt]
Copyright \copyright\ (2024) Silvia Chiacchiera, Patrick B. Warren, Andrew J. Masters, Michael A. Seaton.\\[3pt]
This article is distributed under a Creative Commons Attribution (CC BY) License.

\end{abstract}

\maketitle

\section{Introduction}\label{sec:intro}
\begin{figure}[b]
  \begin{center}
    \includegraphics[width=3.2in]{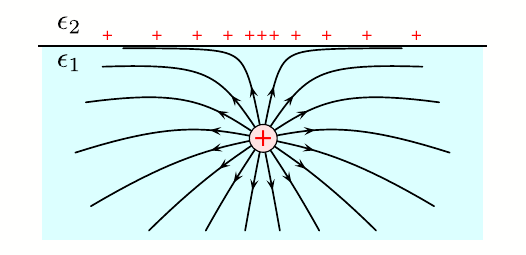}
  \end{center}
  \caption{\small Field lines and induced surface charges for a point charge near an air-water or oil-water interface ($\epssub1 \gg\epssub2$).\label{fig:imagecharge}}
\end{figure}

The relative dielectric permittivity of water ($\epsr\approx78$) is much higher than that of a typical apolar liquid such as a hydrocarbon oil ($\epsb\approx2$).  This means that aqueous structured liquids can have significant dielectric contrasts between water-rich and water-poor regions, and this may modify the distribution of charged species.  Examples of low dielectric regions include the oily cores of surfactant micelles, oil-rich regions in microemulsions or lyotropic liquid crystal phases, the interiors of lipid bilayers, and the oily centres of globular proteins.  In addition, interfaces in such systems (as indeed in simple liquids) present dielectric discontinuities which can influence the local structure, affect the interface properties, and contribute to specific adsorption effects.  For example, repulsion of ions from the air-water interface (\Fig{fig:imagecharge}) leads to ion desorption and a consequent increase in the air-water surface tension \cite{OS34, BZvR08}.

Coarse-grained molecular dynamics methods, such as dissipative particle dynamics (DPD) \cite{FS02, EW17}, are often used to model the properties of structured liquids, and different approaches have been taken to incorporate dielectric effects.  The simplest is to assume a \emph{static} background dielectric permittivity representative of the system as a whole, and capture the effects of local dielectric inhomogeneities in the coarse-grained interaction potentials following a systematic top-down parametrisation strategy \cite{FvM+16, ABF+17_abbr}. This does appear to have some success, for example in modelling surfactant self assembly \cite{VLN13, LVN13, JSJ+16_abbr, ABDR+18_abbr}. However the question remains whether essentially many-body dielectric effects can be truly captured by what are often pairwise coarse-grained interaction potentials.  Further, the transferability of these top-down parameter sets may be limited to systems with similar microstructural motifs.

A more sophisticated approach is an \emph{implicit} method which solves the Poisson equation with an inhomogeneous dielectric matched to the local composition.  For example Groot \cite{Gro03, Gro03b} introduced a particle-particle particle-mesh (\PthreeM) method, with an underlying grid onto which the local dielectric permittivity is mapped.  Implicit methods like Groot's \PthreeM\ typically require bespoke numerical codes, precluding the use of standard methods such as Ewald summation.  But in addition, as Groot remarks, in an implicit dielectric method there must also be forces on the \emph{neutral} solvent particles.

To see why this is the case, focus on \Fig{fig:imagecharge} where a test charge $Q$ is embedded in a medium with a high dielectric permittivity (\eg\ water) at a distance $h$ from an interface with a medium with a low dielectric permittivity (\eg\ air, or oil).  This textbook problem is readily solved to find the charge is repelled from the interface with a force $QQ'/(16\pi\epssub1\epssub0h^2)$, where $Q'=Q(\epssub1-\epssub2)/(\epssub1+\epssub2)$ (the image charge) is located a distance $h$ on the opposite side of the interface.  Of course, the image charge is just a convenient mathematical fiction introduced to solve the inhomogeneous electrostatic problem, and in reality the embedded charge is repelled by induced surface charges as shown in \Fig{fig:imagecharge}.

An implicit dielectric approach will certainly correctly calculate the force on the embedded charge due to the induced charges, but unless one takes into account the induced forces on solvent particles, the \emph{back reaction} on the medium will be missed.  Neglecting this leads to a situation in which there are unbalanced forces, with potentially disastrous consequences for the underlying physics.  Other examples make the same point.  For instance a well-known scientific parlour trick is to use a plastic comb charged by friction to pick up small bits of paper \cite{feyn63}.  This works because the bits of paper, although uncharged, are polarised by the inhomogeneous electric field of the static charges on the comb, and as such feel a force in this field.  Now consider this from the point of view of an implicit dielectric simulation method where the dielectric body comprises solely neutral particles.  If there are no induced forces on these particles, there is no force on the dielectric body in an inhomogeneous field, and a basic physical principle is violated.  To counter this Groot argued the forces on neutral particles (such as solvent beads) can be neglected if spatial inhomogeneities are weak \cite{Gro03}.  We agree, but by the same token the forces on the explicit charges (\eg\ solute ions) due to induced image charges are similarly weak, and can also be neglected.  If this is the case, there is no need to worry about spatial dielectric inhomogeneities in the first place.

Since computing the back-reaction forces in an implicit method is quite onerous, this leads us to explore a third \emph{explicit} approach to modelling dielectric inhomogeneities, one in which we allow the dielectric properties to emerge naturally as a consequence of having explicitly polarisable solvent molecules in the model \cite{Wer79, SPH81, GGJ11}.  In the models we shall explore below, the polarisability arises from partial charges on the (net electrically neutral) solvent molecules.  In this approach, the static background dielectric permittivity is set to a constant (\eg\ representative of a hydrocarbon oil phase) and dielectric inhomogeneities emerge spontaneously corresponding to the distribution of the polar solvent molecules.  Induced charges are explicitly represented, by the disposition of the partial charges. The partial charges interact with other charges, such as the test charge in \Fig{fig:imagecharge}, through the normal Coulomb law.  Force balance is satisfied at all times, and basic physical principles are fully respected.

The penalty introduced here is the need to solve the electrostatics problem including all the partial charges of the solvent molecules. These are of course usually much more numerous that any explicit charges thus there is a considerable additional computational cost. However, this cost replaces the laborious and bespoke calculation of the reaction forces on neutral solvent particles that we have just argued is required in an implicit method.  On the other hand an immediate and clear advantage is that a well-stocked cabinet of molecular dynamics (MD) methods is available to deal with the electrostatics problem \cite{AT87, FS02}.  Mesh discretisation artefacts such as might be encountered using grid-based implicit methods are also obviously absent, but there is no free lunch: other artefacts may arise due to the finite size of the polar molecules, unwanted dielectric saturation, or an unphysical frequency response.  Therefore a systematic approach is required.

Our aim here is to explore the properties of a broad class of polarisable soft solvent models in the paradigm just described, and in the context of dissipative particle dynamics (DPD) as a widely-used prototypical coarse-grained MD method \cite{EW17}.  In the rest of this paper we first describe the models and analysis methods, before reporting on the basic properties in terms of dielectric behaviour. We propose a couple of specific models which could be used for oil-water mixtures, and demonstrate the behaviour in test cases of increasing complexity.

\begin{figure}
  \begin{center}
    \includegraphics[width=3.2in]{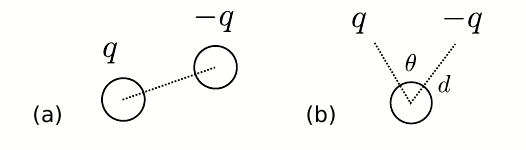}
  \end{center}
  \caption{Polarisable soft solvent models: (a) molecular dimer comprising a bound pair of solvent beads carrying equal and opposite partial charges, (b) trimer comprising a single solvent bead dressed with a pair of tethered partial charges.\label{fig:models}}
\end{figure}

\section{Models}\label{sec:models}
\subsection{Design principles}\label{subsec:design}
In this work polarisable soft solvent models are built by adding partial charges to small solvent `molecules'.  In constructing these models we shall attempt to minimally perturb the `standard' DPD solvent model, which has seen considerable service underpinning various recent systematic parameterisation efforts \cite{FvM+16, ABF+17_abbr}.  This basic underlying model has solvent beads interacting with the standard short-range, pairwise, soft repulsions described by the pairwise interaction potential $\beta\phi(r)=A(1-r/\rc)^2/2$ for $r\le\rc$, and $\phi(r)=0$ for $r>\rc$, where $r$ is the bead-bead centre separation, $A$ is the repulsion amplitude, $\rc$ is a cut-off distance, and $\beta=1/\kT$ is the inverse of the unit of thermal energy $\kT$.  In what follows we shall use $\rc$ and $\kT$ for the simulation units of length and energy, and make the usual choice $\rho\rc^3=3$ for the bead density and $A=25$ for the repulsion amplitude \cite{GW97}.

Building on this, we consider two classes of polarisable solvent model (\Fig{fig:models}).  In the \emph{dimer class} (\partFig{fig:models}{a}), pairs of DPD solvent beads are joined by springs and assigned equal and opposite partial charges $\pm q$. Given the standard choice of bead density, the molecular density in this class is $\rhom\rc^3={3}/{2}$.  In the \emph{dressed solvent class} (\partFig{fig:models}{b}), we shall keep the original solvent beads but to each solvent bead tether a pair of equal and opposite partial charges $\pm q$ by springs.  In this class of trimer models the central beads interact only \via\ the DPD soft repulsions, and partial charges only \via\ electrostatics, thus severing the springs decouples the two systems.  The molecular density in this case is $\rhom\rc^3=3$.

The dimer class has a computational advantage since the density of partial charges is half that of the dressed solvent class, and is favourable if one is only interested in permittivity.  However, a potential disadvantage is that the springs may intrude upon the solvent rheology.  Additionally, Gaussian springs can stretch indefinitely under some flow conditions and FENE springs might be a better choice.  Thus, apart from computational efficiency, one might disfavour this choice.

The dressed solvent class includes models initially investigated by Peter and Pivkin \cite{PP14} and Peter, Lykov and Pivkin \cite{PLP15}, and refined recently by Vaiwala, Jadhav and Thaokar \cite{VJT18}. These models can be regarded as an evolution of the Drude oscillator water model developed for the MARTINI force field \cite{YSS+10,Souza2021_abbr}. The solvent dynamics in the dressed solvent class may also differ from that of the standard DPD solvent, but only indirectly via coupling to the electrostatics.

For both classes, the springs correspond to a bonding interaction potential given by $\beta\phi(d)=\half \kd(d-d_0)^2$ where $d$ is the separation between the force centres, $\kd$ is a spring constant, and $d_0$ the nominal bond length (which may be set to zero).  In addition for the dressed solvent class of models we can allow for an angular spring $\beta\phi(\theta)=\half k_\theta(\theta-\theta_0)^2$ where $\theta$ is the angle subtended at the central bead by the partial charges, $k_\theta$ the angular spring constant, and $\theta_0$ the nominal opening angle.

\subsection{Length scale mapping}\label{subsec:mapping}
The partial charges $\pm q$ are expressed in terms of the fundamental unit of charge, but before proceeding to this step it is first necessary to map $\rc$ into physical units.  To do this, we follow the seminal lead of Groot and Rabone \cite{GR01} in introducing a solvent bead mapping number $\Nmap$ (usually a small integer, but not necessarily so), which is the mean number of real solvent molecules represented by one DPD solvent bead.  It follows that one mole of DPD `volume elements' occupies a volume
\begin{equation}
  \NA\rc^3=\rho\rc^3\Nmap\Vm\,,\label{eq:narc3}
\end{equation}
where $\Vm$ is the solvent molar volume and $\NA$ is Avogadro's number; from this $\rc$ can be calculated.  For example, with the common choice $\Nmap=3$ to represent water ($\Vm\approx0.018\,\mathrm{L}\,\mathrm{mol}^{-1}$), we find $\NA\rc^3\approx0.162\,\mathrm{L}\,\mathrm{mol}^{-1}$ assuming the solvent bead density $\rho\rc^3=3$ as above, which corresponds to $r_c \approx 0.646\,\nm$.  The identification of $\rc$ underpins the conversion of all lengths and molecular densities.  We should emphasise that we do not expect the solvent molecular density and molecular polarisability in the coarse-grained model to be the same as those of the real system.  Rather, the intent is that the solvent model should appear as a featureless dielectric continuum on length scales greater than $\rc$.  Any residual effects of solvent granularity should be viewed as discretisation artefacts.

Throughout the rest of the paper we use DPD units, which amount to setting $k_BT=1$, $m=1$ and $r_c=1$ (that is, in using the thermal energy, DPD bead mass and radius as fundamental units). To help the reader in better grasping the corresponding physical scales, \Table{tab:units} gives the conversion for some key physical quantities, and the values of the reference quantities in SI units for our typical DPD mapping choice.

\begin{table}
  \begin{ruledtabular}
    \begin{tabular}{llll}
      \textrm{Physical quantity}  & DPD unit &
      Example value & unit  \\
      \hline\\[-9pt]
      length & $\rc$  & $0.646$ & $\nm$  \\
      electric field  & $\kT/(e \rc)$
      & $3.98 \times 10^7$ & $\mathrm{V}\,\mathrm{m}^{-1}$\\
      electric dipole & $e \rc$ & $1.03\times10^{-28}$
      & $\mathrm{C}\,\mathrm{m}$ \\
      & & $=31.0$ & D (Debye) \\
      pressure & $\kT/\rc^3$ & 15.3 & MPa \\
      surface tension & $\kT/\rc^2$ & $9.9$
      & $\mathrm{mN}\,\mathrm{m}^{-1}$\\
      kinetic time scale & $\rc \sqrt{{m}/{\kT}}$ & $3.01$ & ps \\[-1pt]
    \end{tabular}
  \end{ruledtabular}
  \caption{DPD units and physical units for relevant physical quantities. The values in the third column correspond to the choice $\Nmap=3$, $\rho\rc^3=3$ and room temperature ($T=298\,\mathrm{K}$).\label{tab:units}}
\end{table}

\subsection{Electrostatics}\label{subsec:electrostatics}
Now we turn to the specification of the electrostatics.  In terms of the dielectric properties, we draw a careful distinction between the following quantities \cite{subnote}:
\begin{equation}
  \begin{split}
    &\epssub0 = \text{absolute permittivity of free space (vacuum)},\\
    &\epsb = \text{static background relative permittivity},\\
    &\epsr = \text{polarisable solvent relative permittivity}.
  \end{split}
\end{equation}
The latter two are defined relative to free space (vacuum).  We further define, relative to background,
\begin{equation}
  \epsilon = {\epsr}/{\epsb}\,.
\end{equation}
The full permittivity of our polar solvent model for example is thus $\epsilon \epsb \epssub0$.

The central problem addressed in the present work is to design a model which achieves a desired value of $\epsilon$.  Thus for example, for water in coexistence with water vapour at room temperature, we should choose $\epsb=1$, and aim for $\epsilon\approx 78$.  We describe such pure water models in more detail in \Sec{sec:waterinvacuum}.  For another example, in a water-in-oil system  we can choose for example $\epsb=2$ for the oil \cite{oilnote}, and then aim for $\epsilon\approx 39$ so that $\epsr=\epsilon\epsb\approx78$ for the water.  We note that in principle one could always use vacuum as a background and make the oil polarisable too, but this would increase the number of charges required in the simulation, since both fluids (oil and water) would comprise polar(isable) molecules, and consequently increase the computational time.

The long-range Coulomb law between unit charges in the \emph{static} background (\eg\ oil) fixes the Bjerrum length,
\begin{equation}
  \lB=\frac{e^2}{4\pi\epsb\epssub0\kT}\,.\label{eq:lb}
\end{equation}
In practical terms, it is the dimensionless ratio $\lB/\rc$ that should be specified, or the dimensionless coupling parameter $\Gamma=4\pi\lB/\rc$.  For instance if the static background is intended to be a hydrocarbon oil we might designate $\epsb=2$ for which $\lB\approx 27\,\nm$, and then $\lB/\rc\approx 42$ ($\Gamma\approx 530$), assuming the $\Nmap=3$ mapping.  For another example, for vacuum as background one would have $\lB\approx55\,\nm$ and $\lB/\rc\approx 85$ ($\Gamma\approx1070$).

The final part of the electrostatics specification concerns charge smearing, which is required to prevent a `collapse' since there are no hard cores \cite{MEF66}.  Here we consider both Gaussian charge smearing \cite{WVA+13, WV14}, and Slater charge smearing \cite{GMV+06, IMT09, VJT17}.  For Gaussian charges the interaction potential (between partial charges) is
\begin{equation}
  \beta\phi(r)=\frac{\lB q^2}{r}
\erf\Bigl(\frac{r}{2\sigma}\Bigr)\,,
\label{eq:dpdvlr}
\end{equation}
where $\sigma$ is the smearing length.  For Slater partial charges (which are used in the models described in \Sec{sec:waterinvacuum}) we use the approximate form
\begin{equation}
  \beta\phi(r)=\frac{\lB q^2}{r}
  [1-(1+\betastar\rstar)e^{-2\betastar\rstar}]\,,
  \label{eq:gm}
\end{equation}
where $\betastar=\rc/\lambda$, $\rstar=r/\rc$, and $\lambda$ is the Slater smearing length.  In both of these the correct long-range Coulomb law is recovered for large separations.

To summarise: the polarisable solvent models we study are specified by the molecular parameters $\kd$, $d_0$, and additionally for trimers $k_\theta$, $\theta_0$\,; the magnitude of the partial charges $q$\,; the charge smearing scheme and length $\sigma$ or $\lambda$\,; the choice of static background $\epsb$ which sets the Bjerrum length $\lB$\,; the repulsion amplitude between solvent beads $A$\,; the bead or molecular density $\rhom$\,; and the $\Nmap$ mapping choice which fixes $\rc$.  The key target is the ratio $\epsilon=\epsr/\epsb$.

\subsection{Specific models}\label{subsec:models}
Before listing the specific models used in the present work, let us make some general remarks.  As mentioned, the central aim of a \emph{mesoscale} polarisable solvent model is to provide a featureless dielectric continuum, in so far as the solvent beads are uniformly distributed.  To achieve this we should endeavour to keep structural features and artefacts to length scales less than $\rc$. Following this line of argument, in order to get a large value of $\epsilon$ we should make the individual molecular dipoles as large as possible, but with a bond length not larger than $\rc$.  This thinking drives the essential choices below.  Nevertheless, it shall become apparent that there are dipole-dipole correlations in these models, which obliges care in applications.

In the \emph{dimer} (DIM) class, we fix $\rhom r_c^3=3/2$, $A=25$, $k_d=5$, $d_0=0$, and use Gaussian charge smearing with $\sigma=0.5$. Then $l_B$ and $q$ are varied to match the target system.  In particular, for water in oil we suggest $(l_B/r_c,q)=(42,0.463)$, and call the parameterisation \WinoDim.

In the \emph{dressed solvent} (\DS) class, we fix $\rhom r_c^3=3$, $A=25$, $k_d=10$, $d_0=0$, and use Gaussian smearing with $\sigma=0.5$. Here, for water in oil we suggest $(l_B/r_c,q)=(42,0.36)$, and call the parameterisation \WinODS. Within the same class, five other models obtained by varying  the partial charge are considered (namely: $l_B/r_c=42$ and $q = 0.08, 0.1, 0.2, 0.3, 0.4$).  We call these parameterisations \DSq{0.08}, \DSq{0.1}, and so on.  The dielectric properties are reported in \Sec{subsec:epsres}, and for practical reasons some are used in the tests of \Sec{subsec:tests}.

In the codes the long range electrostatics part is dealt with by Ewald summation, as discussed in \Appendix{app:ewald}, or in the case of \DLMESO\ optionally also by smooth particle mesh Ewald (SPME) \cite{Essmann1995}. In physical terms, neither $\sigma$ nor $\lambda$ should have any significance, and the actual choice is dictated by pragmatic considerations \cite{WVA+13, WV14}.  For completeness, the parameterisations \WinoDim\ and \WinODS\ are reported in \Table{tab:dimtri}.

A subset of the dressed solvent class includes the Peter-Pivkin water-in-vacuum models \cite{PP14,PLP15} with strong harmonic bonds and the additional angle potential. Here we would like to highlight that they do not lead to the desired permittivity for water without modification. We discuss this class of model, including parameterisation challenges, more fully in \Sec{sec:waterinvacuum}.

\begin{table}
  \begin{ruledtabular}
    \begin{tabular}{lll}
      model class $\rightarrow$  & dimer solvent & dressed solvent \\
      \hline\\[-9pt]
      parametrisation &  \WinoDim\  & \WinODS\  \\
      \hline\\[-9pt]
      $r_c$ & $0.646\,\nm$ & $0.646\,\nm$ \\
      $A$ & 25 & 25\\
      $\kd/\rc^{-2}$ & 5 & 10\\
      $q$ & 0.463 & 0.36\\
      $\sigma/\rc$ & 0.5 & 0.5  \\
      $\lB$ & $27\,\nm$ & $27\,\nm$ \\
      $\lB/\rc$ & 42 &  42\\
      $\rhom\rc^3$ & 3/2 & 3 \\[3pt]
      \hline\\[-9pt]
      $\myav{\drvec^2}/\rc^2$ & 0.669(1) & 0.5660(5) \\
      $\gC$ & 1.02(4) & 1.05(5) \\
      $\epsr/\epsb=\epsilon$ & 40(2) & 42(2)\\
      $\gK$ & 0.69(3) & 0.70(2) \\
      \hline\\[-9pt]
      $\epsb$ (oil) & 2 & 2 \\
      $\epsr$ (water) & 80(4) & 84(4)\\[-1pt]
    \end{tabular}
  \end{ruledtabular}
  \caption{Parameters of the dimer solvent and dressed solvent models, with the values of $q$ and $\lB$ appropriate for water in oil. As explained in the text, the standard DPD soft repulsion acts between all beads in the dimer class, whereas only between neutral beads in the dressed solvent class. The dielectric properties (see \Sec{subsec:electrostatics}, \ref{subsec:epsr} and \ref{subsec:corr}) are computed here from Monte-Carlo simulations (see also \Table{tab:dsprop}).  A figure in brackets here and elsewhere is an estimate of the error in the final digit.\label{tab:dimtri}}
\end{table}

\section{Methods}\label{sec:methods}
\subsection{Simulations}\label{subsec:codes}
We study the properties of the models introduced above with a combination of dissipative particle dynamics (DPD) using \DLMESO\ \cite{SAM+13} \cite{ECAM}, and Monte-Carlo (MC) simulations using a bespoke code.  Whereas \DLMESO\ can be applied to a wide variety of interesting problems, the MC code is limited to generating homogeneous solvent configurations to compute structural and thermodynamic solvent properties, including of course the relative permittivity $\epsilon=\epsr/\epsb$.

For the DPD simulations, unless otherwise stated, the typical simulated volume is a cubic box of side $L=10$. The time step is always $\dt=0.01$, and simulations are typically run for $5\times 10^5$ time steps, after $10^4$ equilibration steps. Sampling is done every 100 steps. The DPD drag coefficient is $\gamma=4$.

Whilst \DLMESO\ is well documented, we here provide some brief details of the MC code, which was based on earlier work \cite{WVA+13}. For present purposes this was configured to run in an $NVT$ ensemble, with single-particle trial displacements.  A standard Metropolis scheme is implemented \cite{FS02}, with trial displacements $\delta r=0.25$--0.4 chosen to obtain an acceptance rate of 30--50\%.  We typically consider cubic boxes of side $L=8$, and typically equilibrate for $3\times10^5$ trial MC moves before sampling the configuration every $1.5\times10^5$ trial moves.  One can show that this is sufficient to allow all the particles to move a distance of order 1--$2\,\rc$ between sampled configurations.  A large number (500--1000) of samples are required to estimate the mean square box dipole moment with sufficient accuracy.  We achieve this in part by task-farming across a 100+ node cluster (at the expense of having to equilibrate on each node).

Benchmarking \DLMESO\ against the MC code allows us to test for issues due to incomplete equilibration, as well as possible artefacts coming from the DPD thermostat (which, recall, is a pairwise spring-dashpot type); no significant effects were found.  Also, the independently developed MC code uses interparticle potentials, whereas \DLMESO\ uses interparticle forces, which provides a stringent test for coding errors in the implementation of the electrostatics, and in fact in the course of the work we uncovered a problem with the implementation of Slater smearing in \DLMESO\ which can be traced to an incorrect expression for the forces in the literature \cite{IMT09}; it was obviously fixed for the present study.  For completeness, in \Appendix{app:ewald} we document the nature of the problem and provide some precision benchmark MC results for the reference Slater charge plasma (see below).

To summarise, \DLMESO\ and the bespoke MC code are found to be in excellent mutual agreement.  This gives us a high degree of confidence in the results.

\subsection{Liquid state theory}\label{sec:hnc}
For supporting calculations we shall occasionally use a liquid state theory to calculate the structural properties of multicomponent charged fluids (plasmas).  This was developed for our previous work \cite{WVA+13, WV14}, and uses the hypernetted chain (HNC) approximation to close the Ornstein-Zernike equations \cite{HM06}.  The HNC is known to yield particularly accurate results for soft potentials \cite{FvM+16}, and  takes proper account of the electrostatics such as the Stillinger-Lovett sum rules \cite{HM06}.  Some benchmark results related to the present problem are included in \Table{tab:benchmark} in \Appendix{app:ewald}.

\subsection{Dielectric permittivity}\label{subsec:epsr}
To calculate the relative dielectric permittivity, we use either \DLMESO\ or MC to generate configurations of $\Nm$ solvent molecules in a cubic simulation box of volume $L^3=V$.  Then, in linear response theory, the dielectric constant can be computed from fluctuations in the box dipole moment as described in Frenkel and Smit \cite{FS02}, Allen and Tildesley \cite{AT87}, and Kusalik \etal\ \cite{KMS94}.  Note that there is scope for considerable confusion over units here, with some authors preferring Gaussian (cgs) units, and others using bespoke reduced units.  In the present work we shall formulate the problems initially using SI units, but switch to writing expressions in terms of the Bjerrum length $\lB$ where practically possible.

For the most part we adopt the so-called `tin-foil' or \emph{conducting} boundary conditions since this gives the most accurate results for a given computational cost \cite{AT87}.  To summarise the calculation in the present context, let the dipole of the $i$-th solvent molecule be $\pvec_i$ and the total dipole moment in the simulation box be $\Pbox = \sum_{i=1}^{\Nm} \pvec_i$ \cite{boxnote}.  Then \cite{AT87, FS02}
\begin{equation}
  \epsilon
  =1+\frac{\myavC{\Pbox^2}}{3\epsb\epssub0V\kT}\,,
  \label{eq:er1}
\end{equation}
where, to recall, $\epsilon={\epsr}/{\epsb}$ is the permittivity of the polarisable solvent relative to the static background, and $\myavC{\dots}$ denotes an ensemble average.  Note that the static background relative permittivity features also in the denominator on the right-hand side of this equation.

In terms of the Bjerrum length defined in \Eq{eq:lb}, the above expression can be written as
\begin{equation}
  \epsilon = 1+3y\gC\,,\label{eq:cond}
\end{equation}
where $y = {4\pi\lB\rhom\myav{\pvec^2}} / {9e^2}$ and $\gC = {\myav{\Pbox^2}} / {\Nm\myav{\pvec^2}}$, following an established notation \cite{AT87, HM06}.  Here $\rhom = \Nm / V$ is the number density of solvent molecules, $\myav{\pvec^2}$ is the mean square molecular dipole moment, and the subscript `C' on $\gC$ signals the choice of (conducting) boundary conditions.

The solvent molecules in our models have equal and opposite partial charges, so in practical terms $\pvec_i=q e \drvec_i$ where $\drvec_i$ is the vector from $-q$ to $q$.  We additionally define $\DRvec=\sum_{i=1}^N \drvec_i$.   For charge smeared models, the charge clouds are centro-symmetric and $\drvec_i$ is measured between the charge cloud centres.  From these,
\begin{equation}
  y = \frac{4\pi\lB q^2\rhom\myav{\drvec^2}}{9}\,,
  \quad\gC = \frac{\myavC{\DRvec^2}}{\Nm\myav{\drvec^2}}\,.\label{eq:ygC}
\end{equation}
The values of $\myav{\drvec^2}$ and $\myavC{\DRvec^2}$ required in these expressions (in units of $\rc^2$) can be obtained from a time series of simulation snapshots. Care is taken to ensure the simulation snapshots are uncorrelated, by measuring the time autocorrelation function for $\DRvec$.  Errors are estimated by block averaging, and further controlled by replicate simulation runs.

\begin{figure}
  \begin{center}
    \includegraphics[width=3.2in]{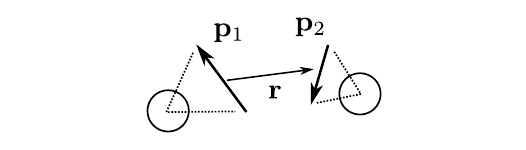}
  \end{center}
  \caption{Geometry for molecular dipole distribution functions: for each pair of solvent dipoles (here illustrated for the trimer model) we define the dipole moments $\pvec_1$ and $\pvec_2$, and the distance between the dipole centers $\rvec$.\label{fig:structure}}
\end{figure}

\begin{table*}
  \begin{ruledtabular}
    \begin{tabular}{llllllllll} 
      \multicolumn{2}{l}{Parametrisation} & \DSq{0.08} &  \DSq{0.1} &
      \DSq{0.2} &  \DSq{0.3} & \multicolumn{2}{l}{\WinODS\ (or \DSq{0.36})}  & \DSq{0.4} \\[1pt]
      \hline\\[-9pt]
      Method & Quantity & $q=0.08$ & $q=0.1$ & $q=0.2$ & $q=0.3$ & $q=0.36$ & [$\epsilon'=42$] & $q=0.4$ \\
      \hline\\[-9pt]
      DPD & $\myav{\drvec^2}/\rc^2$ & 0.5926(5) & 0.5898(3) & 0.5785(3) & 0.5709(3) &  0.5669(3) & & 0.5648(3) \\
          & $\gC$                  & 1.00(2)  & 1.02(3) & 1.08(3)  & 1.08(3)   & 1.07(3)   & & 1.00(4) \\
          & $\epsr/\epsb=\epsilon$ & 2.99(5)  & 4.18(9) & 14.2(4) & 30.3(8) & 42(1) & &  49(2) \\
          & $\gK$                  & 0.78(1)  & 0.76(2) & 0.75(2)  & 0.73(2)   & 0.72(2)   & & 0.67(3) \\
      \hline\\[-9pt]
      Monte-Carlo & $\myav{\drvec^2}/\rc^2$ & 0.593(1) & 0.590(1) & 0.579(1) & 0.571(1) &  0.5660(5) & 0.5660(4) & 0.564(1) \\
      & $\gC$                     & 1.03(6) & 1.05(6) & 1.01(6)  & 1.02(3)    & 1.04(4)   &           & 1.09(3) \\
      & $\epsr/\epsb=\epsilon$    & 3.1(1) & 4.3(2) & 13.4(7) & 29(2) & 41(2) & 43(2) & 53(2) \\
      & $\gK$                     & 0.80(4) & 0.78(4) & 0.70(4)  & 0.69(2)    & 0.70(2)   & 0.72(2)   & 0.73(2) \\
      pressure & $\rc^3\press/\kT$ & 23.64(2) & 23.62(2) & 23.50(2) & 23.37(2) & 23.26(1) & & 23.21(2)\\
      \hline\\[-9pt]
      Wertheim & $\myav{\drvec^2}/\rc^2$ & 0.5948 & 0.5934 & 0.5874
      & 0.5830 & 0.5809 & & 0.5796 \\
      Onsager & $\epsr/\epsb$ & 3.65 & 5.29 & 19.1 & 42.1 & 60.1 & & 73.9\\[-1pt]
    \end{tabular}
  \end{ruledtabular}
  \caption{Tabulated results for all dressed solvent models as a function of partial charge.  Results are shown for DPD using \DLMESO\ and for comparison from Monte-Carlo simulations, for which we also report the pressure $\press$.  Generally, the first two rows are computed directly from simulation, being the mean square molecular dipole length $\myav{\drvec^2}$, and the factor $\gC=\myavC{\DRvec^2}/\Nm\myav{\drvec^2}$ computed under conducting boundary conditions.  These are used to calculate the relative permittivity $\epsr/\epsb$ (third row) from \Eqs{eq:cond} and~\eqref{eq:ygC}, and the Kirkwood factor $\gK$ (fourth row) from \Eq{eq:fromgC}.  The penultimate row labelled `Wertheim' is from the theory in \Sec{subsec:lstheory} with the plasma reference correction; without this the entries would all be $\myav{\drvec^2}/\rc^2=3/5=0.6$.  The final row utilises Onsager theory to estimate the relative permittivity from \Eq{eq:ons2}, which assumes $\gK=1$.  For the \WinODS\ model ($q=0.36$), an extra column is inserted for properties computed in Monte-Carlo with a reaction field, where the embedding medium relative permittivity is set to $\epsilon'=42$ to match that of the fluid (or, close enough for present purposes).\label{tab:dsprop}}
\end{table*}

\subsection{Correlation functions}\label{subsec:corr}
To quantify the structural properties we examine the solvent bead pair distribution function, $g_{00}(r)$, and the molecular dipole-dipole correlation functions following Stell \etal\ \cite{SPH81}.  For the latter we introduce a number of radially-resolved moment correlation functions (see \Fig{fig:structure} for the geometry),
\begin{equation}
\begin{array}{l}
g_{000}(r)=\myrav{1}\,,\quad
g_{110}(r)=\myrav{\pvec_1\cdot\pvec_2}\,,\\[3pt]
g_{112}(r)=\myrav{\frac{3}{2}
(\pvec_1\cdot\hat{\rvec})
(\pvec_2\cdot\hat{\rvec})-\frac{1}{2}
\pvec_1\cdot\pvec_2}\,,
\end{array}\label{eq:gnnn}
\end{equation}
where $r=|\rvec|$, $\hat{\rvec}=\rvec/r$, and $\myrav{\dots}$ indicates a radially-resolved average.  For normalisation we ensure $g_{000}(r)\to1$ as $r\to\infty$ so that it behaves like a standard pair distribution function, and for the higher moments we ensure that $\int \dtr\, g_{110}(r)\equiv (1/\Nm)\sum_{i\ne j}\myav{\pvec_i\cdot\pvec_j}$\,, and \mutmut\ for $g_{112}$.

These correlation functions (involving dipole moments) should be measured under \emph{matched} or `Kirkwood' boundary conditions in which the dielectric permittivity of the embedding medium is the same as that of the system itself, otherwise there are spurious artefacts on the length scale of the simulation box \cite{g110note}.  To simulate under these conditions one needs to include, in addition to the inter-particle potential, a `reaction field' term \cite{AT87, FS02}, which in terms of the Bjerrum length is
\begin{equation}
  \beta\phipol = \frac{2\pi\lB \Pbox^2}{(2\epsilon'+1)Ve^2}\,.\label{eq:vpol}
\end{equation}
The permittivity $\epsilon'$ in here is that of the embedding medium (relative to background), which should be matched to that of the fluid (\ie\ $\epsilon'=\epsilon$). This means that the relative permittivity $\epsilon$ of the fluid should be pre-computed, for instance using conducting boundary conditions as described in the preceeding section.

Under these matched ($\epsilon'=\epsilon$) boundary conditions, the expression in \Eq{eq:cond} for the dielectric permittivity is replaced by the Onsager-Kirkwood expression \cite{AT87, FS02, Boettcher73},
\begin{equation}
  \frac{(\epsilon-1)(2\epsilon+1)}{9\epsilon}=y\gK\,.\label{eq:ons}
\end{equation}
Here the `Kirkwood factor' $\gK={\myavK{\Pbox^2}}/{\Nm\myav{\pvec^2}}= \myavK{\DRvec^2}/\Nm\myav{\drvec^2} < \gC$\,, since $\phipol$ acts to suppress fluctuations in the box dipole moment, making $\myavK{\Pbox^2}$ smaller than under conducting boundary conditions.  On the other hand,  to $O(1/\Nm)$, $\myav{\pvec^2}$ is unaffected by the choice of boundary conditions \cite{GGJ11}.

Since $\myavC{\Pbox^2} = \Nm\myav{\pvec^2} + \sum_{i\ne j}\myav{\pvec_i\cdot\pvec_j}$\,, the Kirkwood factor can also be computed from $g_{110}(r)$ as
\begin{equation}
  \gK=1+\frac{1}{\myav{\pvec^2}}
  \int_0^\infty\!\!\dr\; 4\pi r^2\,g_{110}(r)\,.\label{eq:pint}
\end{equation}
Thus $g_{110}(r)$ quantifies the dipole-dipole correlations which contribute to $\gK\ne1$.  For completeness, the function $g_{112}(r)$ is included in the list of computed correlation functions, and the corresponding integral is, apart from a constant of proportionality, approximately the energy associated with the dipoles (exactly so, for true point dipoles).

Since $\myav{\pvec^2}$ and therefore $y$, as defined below \Eq{eq:cond}, are unaffected by the choice of boundary conditions, eliminating $y$ between \Eqs{eq:cond} and~\eqref{eq:ons} obtains
\begin{equation}
  \frac{\gK}{\gC}=\frac{2\epsilon+1}{3\epsilon}\,.\label{eq:fromgC}
\end{equation}
This allows us to compute the Kirkwood factor $\gK$, given the value of $\gC$ computed under conducting boundary conditions and the permittivity $\epsilon$.

\begin{figure}
  \begin{center}
    \includegraphics[width=3.2in]{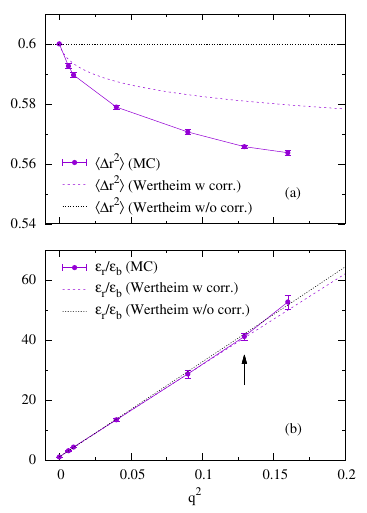}
  \end{center}
  \caption{Dielectric properties of dressed solvent models in \Table{tab:dsprop}, as a function of $q^2$: (a) mean square dipole length $\myav{\drvec^2}$, and (b) relative permittivity $\epsilon=\epsr/\epsb$; the arrowed point is the WinO-DS model (\Table{tab:dimtri}), with $q=0.36$ and $\epsilon=42\pm2$. Markers with error bars are Monte-Carlo simulations (error bars not shown in (a) as the error is smaller than markers). Lines are liquid state theory predictions with (dashed lines) and without (dotted lines) the plasma reference state correction.\label{fig:properties}}
\end{figure}

\subsection{Wertheim and Onsager theories}\label{subsec:lstheory}
A relatively simple liquid state theory can be used to predict the value of $\myav{\drvec^2}$ in the above section, and from there the permittivity itself.  The approach is rigorously based in Wertheim perturbation theory and more details will be presented elsewhere \cite{Mas18}. The first-order theory results in an approximate expression for the probability distribution $\prob(\Dr)$ for the distance $\Dr$ between partial charges, with
\begin{equation}
  \prob(r)= C \exp[-\beta \phibond_{+-}(r)]\>
  \gref_{+-}(r)\,.\label{eq:lst1}
\end{equation}
Here $C$ is such that $4\pi\int_0^\infty\dr\,r^2\,\prob(r) = 1$.  In \Eq{eq:lst1}, $\phibond_{+-}(r)$ is the intramolecular pair potential between the partial charges arising from the bonds and $\gref_{+-}(r)$ is the pair distribution function between the partial charges in a reference system in which all the bonds have been cut.  Given all this,
\begin{equation}
  \myav{\drvec^2}\approx
  4\pi\int_0^\infty\!\!\dr\;r^4\,\prob(r)\,.
       \label{eq:lst2}
\end{equation}
Details of the intramolecular potentials for the models studied in the present work are given in \Appendix{app:intramol}.

In the reference system (\ie\ where the intramolecular bonds have been cut) $\gref_{+-}$ is the pair distribution function between unlike charges in a neutral two-component \emph{plasma}, at a total number density set by the original molecular solvent density ($\rhom$ for the dimer case, $2\rhom$ for the trimer case). In the following sections, we refer to such a reference system as a \emph{reference plasma}. To calculate $\gref_{+-}$ we can therefore deploy the HNC liquid state theory described in \Sec{sec:hnc}. However, since the plasma is generally weakly coupled, one can already get quite accurate predictions by setting $\gref_{+-}=1$.  This gives estimates for $\myav{\drvec^2}$ in some cases in a closed analytic form. In principle higher order terms can be included in the perturbation expansion, but these are harder to calculate \cite{Mas18}.

This now allows us to compute $y$ from the first of \Eqs{eq:ygC}, and from this $\epsilon$ itself can be estimated using Onsager theory \cite{Boettcher73} in the form $(\epsilon-1)(2\epsilon+1) / \epsilon = 9y$, which solves to
\begin{equation}
  \epsilon \approx
  \frac{{1+9y+3\surd({1+2y+9y^2})}}{4}\,.\label{eq:ons2}
\end{equation}
Onsager theory is equivalent to setting $\gK=1$ in (the exact) \Eq{eq:ons}, and so involves the further approximation of the neglect of the dipole-dipole correlations.  The more familiar (but even more approximate) Clausius-Mossotti relation, namely $(\epsilon-1) / (\epsilon+2) = y$, cannot be used if $y>1$ (which is usually the case here) since that lies outside its domain of validity \cite{Boettcher73}.

\begin{figure*}
  \begin{center}
    \includegraphics[width=5.2in]{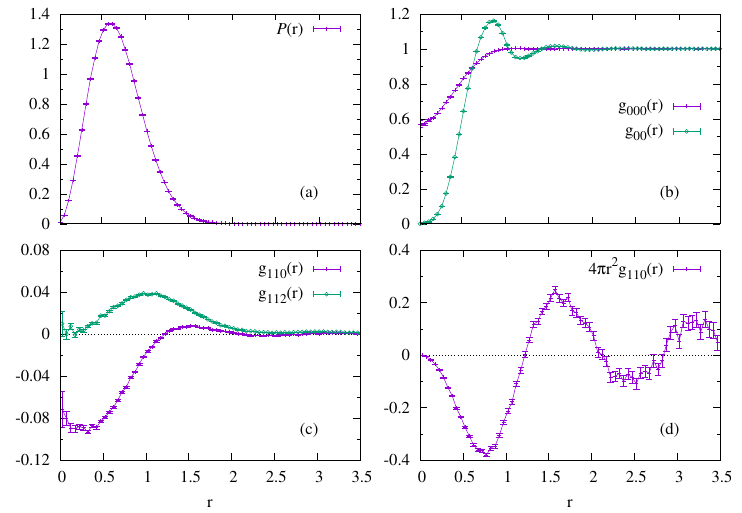}
  \end{center}
  \caption{Correlation functions for the \WinODS\ model (\Table{tab:dimtri}): (a) dipole length distribution, (b) central bead pair distribution function $g_{00}(r)$ and dipole-dipole centre distribution function $g_{000}(r)$, (c) and (d) higher order dipole-dipole correlation functions $g_{110}(r)$ and $g_{112}(r)$. The functions in (c) and (d) are normalised by dividing by $\myav{{\mathbf p}^2}$.  These functions are computed using matched `Kirkwoood' boundary conditions, with $\epsilon'=42$ in the reaction field, as described in \Sec{subsec:corr} (see also \Table{tab:dsprop}).\label{fig:distributions}}
\end{figure*}

\section{Results}\label{se:results}
In this section we report on the dielectric properties of the proposed solvent
models, then describe a series of problems that allow us to test
them.  Concrete examples are given for the class of dressed solvent (DS) model.

\begin{figure}
  \begin{center}
    \includegraphics[width=3.2in]{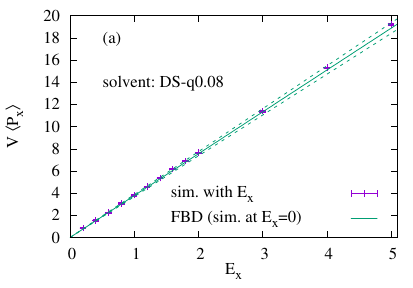}
    \includegraphics[width=3.2in]{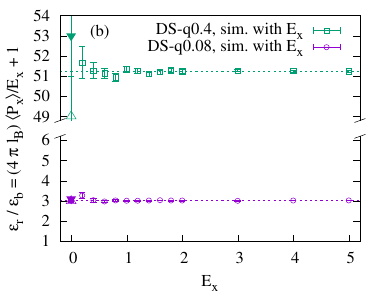}
  \end{center}
  \caption{Application of an external electric field. (a) Mean polarisation under an applied field (points), compared to prediction from \Eq{eq:field} in the text (lines) using $\epsr/\epsb=2.99(5)$ from the fluctuating box dipole method (\Table{tab:dsprop}). (b) Relative permittivity computed from this response as a function of the applied field. Results are from DPD simulations with dressed solvent models having $q=0.08$ and $q=0.4$. The horizontal lines are fits. For comparison, points at $E_x=0$ show the prediction from the fluctuating box dipole method in DPD (open upward triangles) and MC (filled downward triangles).\label{fig:response}}
\end{figure}

\subsection{Dielectric properties}\label{subsec:epsres}
As already described, we have explored the properties of both the dimer solvent and dressed solvent models.  Our results for the dimer solvent are similar to those for the dressed solvents, and so we report only on the latter in detail.  The relative permittivity $\epsilon=\epsr/\epsb$ and the $\gC$ factor are given in the second column in \Table{tab:dimtri} for the water-in-oil dimer model `WinO-DIM', in \Table{tab:dsprop} and \Fig{fig:properties} for the dressed solvent models in general, and in the third column in \Table{tab:dimtri} for the water-in-oil specific model `WinO-DS'. Apart from certain simulations of the WinO-DS model, all these results are obtained using the fluctuating box dipole method with conducting boundary conditions as described in \Sec{subsec:epsr}.  To gain insight into the \role\ of dipole-dipole correlations, the Kirkwood factor $\gK$ is included, calculated from $\epsilon$ and $\gC$ using \Eq{eq:fromgC}.

For the majority, we compute these properties both using the \DLMESO\ DPD code, and separately with our bespoke Monte-Carlo (MC) code. We find excellent agreement between these, which gives us a high degree of confidence in the results.

We also report the MC pressure for the dressed solvent models, observing that it diminishes very slightly with increasing $q$ because of the net cohesive effect of adding the partial charges, but is otherwise very close to the canonical DPD solvent ($\rho\rc^3=3$, $A=25$) for which we find the pressure in reduced units $\rc^3\press/\kT=23.65\pm0.02$.  This suggests that the addition of tethered partial charges in these models is only minimally perturbative to the underlying DPD solvent model.

For the dielectric properties, we can see from \Table{tab:dsprop} and \partFig{fig:properties}{a} that the mean square dipole length  $\myav{\drvec^2}$ is relatively insensitive to the magnitude of the partial charges, although it decreases somewhat with increasing $q$, which is to be expected as opposite charges attract.  For the dressed solvent models, in the absence of a plasma correction $\myav{\drvec^2}=3\kT/(\kd/2)$ by equipartition, where $\kd/2$ is the effective spring constant, as argued in \Appendix{app:intramol}.  This results in $\myav{\drvec^2}=3/5$ since $\kd=10$, which is already close to the reported results.  If the plasma reference correction is additionally included, as in \Sec{subsec:lstheory}, then the agreement between the theory (Wertheim) and the model improves still further, as seen by comparing the corresponding rows in \Table{tab:dsprop}.

By far the biggest controlling factor for the relative dielectric permittivity is therefore the magnitude of $q$, and this is the reason we chose this as the main adjustable parameter in our models.  The dependence of $\epsilon=\epsr/\epsb$ on $q$ is shown in \partFig{fig:properties}{b}, which indicates that $\epsilon$ grows approximately linearly in $q^2$ for $\epsilon\gg1$.  As a design criterion, this observation allows us to relatively easily build models to reproduce targeted values of the dielectric permittivity.  This will be illustrated in relation to the Peter-Pivkin water-in-vacuum models in \Sec{sec:waterinvacuum}.

The Kirkwood factor is roughly constant across all the models, including the dimer model WinO-DIM, and takes the value $\gK\approx0.7$--0.8.  This indicates that dipole-dipole correlations are not negligible in these models.  This is verified by the correlation functions calculated for WinO-DS model, shown in \Fig{fig:distributions}.  Note in particular the damped oscillatory nature of $g_{110}(r)$ whose integral is directly related to $\gK$ by \Eq{eq:pint}.  As mentioned, these correlation functions are computed under matched boundary conditions, by including the appropriate reaction field and utilising the relative permittivity computed under conducting boundary conditions.  This allows also a direct computation of $\gK$ and a separate estimate of the relative permittivity, which are found to be in excellent agreement with those calculated using conducting boundary conditions (see columns labelled WinO-DS in \Table{tab:dsprop}).

Given that the Kirkwood factor $\gK\approx0.7$--0.8, Onsager theory  (final row in \Table{tab:dsprop}) over-predicts the dielectric permittivity by approximately 20--30\%.  Nevertheless, the combination of Wertheim and Onsager allows a useful initial estimate of these properties, before resorting to simulation. 

\subsection{Test problems}\label{subsec:tests}
Here, various physical situations are analyzed to confirm that the solvent behaves as expected, \ie\ as a medium having the dielectric permittivity as determined by the fluctuating box dipole method.  DPD simulations with models belonging to the dressed solvent class are used for this purpose.

In the presence of an electric field, the polarisation density from classical electrostatics is $\Pvec=(\epsr-1)\epssub0\Evec$.  For our polarisable molecular solvent with a \emph{background} relative permittivity, this generalises to the \emph{induced} polarisation of the solvent (\ie\ the response of the partial charges),
\begin{equation}
  \Pvec=(\epsilon-1)\epsb\epssub0\Evec\,,\label{eq:pvec}
\end{equation}
where, as usual and as a reminder, $\epsilon=\epsr/\epsb$.

All three methods below can be reverse-engineered to calculate the ratio $\epsr/\epsb$ and the results are summarised in \Table{tab:tests}, compared to the consensus results from fluctuating box dipole method using DPD and MC.  In general the agreement is very good.  As a practical comment, we note that measuring the induced polarisation with an applied electric field is the most accurate of these.

\begin{table}
  \begin{ruledtabular}
    \begin{tabular}{llcc}
       Parametrisation &  & DS-q0.08 & DS-q0.4 \\
       \hline\\[-9pt]
       Method & Refer to & $q=0.08$ & $q=0.4$ \\
       \hline\\[-9pt]
       Fluctuating box dipole
       & \Table{tab:dsprop}
       & 3.0(1)\phantom{00} & 51(2)\phantom{.00} \\
       Applied field & \Sec{subsubsec:testfield}
       & 3.021(3) & 51.24(1) \\
       Force reduction & \Sec{subsubsec:testforce}
         & 3.03(4)\phantom{0} & 44(7)\phantom{.00} \\ 
       Charge screening & \Sec{subsubsec:testscreen}
        & 3.00(7)\phantom{0}& ---     \\[-1pt] 
    \end{tabular}
  \end{ruledtabular}
  \caption{Values of $\epsr/\epsb$ from various tests described in the main text, compared to the consensus (DPD and MC) values using the fluctuating box dipole method (FBD) reported in \Table{tab:dsprop}, for dressed solvent models with the indicated values of the partial charge $q$.\label{tab:tests}}
\end{table}

\subsubsection{Application of an electric field}\label{subsubsec:testfield}
As a first test, an electric field $E_x$ along the $x$ direction is applied to a simulation box containing the polarisable solvent molecules at the standard density, and the resulting average polarisation is measured. In \partFig{fig:response}{a} this \emph{direct} response (points) is compared with the linear response expected using the permittivity from the fluctuating box dipole method (solid line).

In reduced units, \Eq{eq:pvec} becomes for this problem
\begin{equation}
  \frac{q\Delta R_x \rc^2}{V}=\Bigl(\frac{\epsr}{\epsb}-1\Bigr)\,
  \frac{\rc}{4\pi\lB}\times\frac{e\rc E_x}{\kT}\,.\label{eq:field}
\end{equation}
The quantity on the left-hand side is the polarisation density in the box in reduced units, and the second factor on the right-hand side is the electric field expressed as a force per unit charge, again in reduced units.  Measuring the slope of the response therefore allows us to compute $\epsr/\epsb$, which is shown for dressed solvent models with $q=0.08$ and $q=0.4$ in \partFig{fig:response}{b}, with the final results reported in \Table{tab:tests}.  In general, a good agreement is found, confirming that the dielectric responds linearly to the external field, for the field strengths considered, \ie\ at least up to $5 \kB T/ (e r_c)$.

\begin{figure}
  \begin{center}
    \includegraphics[width=3.2in]{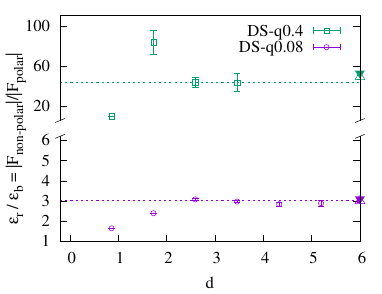}
  \end{center}
  \caption{Force reduction (points) as function of the distance between the charges, $d/r_c$, and its best estimate for $d>2\rc$ (dashed lines). Results are from DPD simulations with test charges $Q=\pm 5$ in dressed solvent models with $q=0.08$ or $q=0.4$. Note that points for $q=0.4$ and $d>4\rc$ are not shown, as $\Fpolar$ is too small to be measured. For comparison, points at large $d$ show the prediction from the fluctuating box dipole method in DPD (open upward triangles) and MC (filled downward triangles).\label{fig:reduction}}
\end{figure}

\subsubsection{Force between two test charges}\label{subsubsec:testforce}
In this second test, two charges $\pm Q$ are fixed at a distance $d$ from each other and embedded in a solvent. Passing from an apolar to a polar medium, the force acting on each one is reduced by a factor $\epsb/\epsr=|\Fpolar|/|\Fnonpolar|$.  Since we work with simulation boxes with periodic boundary conditions, what is actually reported here is the force reduction per charge, averaged over the three axis directions, and both ions, for a pair of interleaved, infinite cubic arrays of equal and opposite charges, separated by a distance $d$ along the diagonal.

In \Fig{fig:reduction} we show results for the force reduction (points) as a function of $d/r_c$. The polar (non-polar) medium is represented here by the dressed solvent model with $q\neq 0$ ($q=0$), and we choose $Q=\pm5$.  For large enough $d$, it can be seen that the force reduction agrees nicely with $\epsr$ from the fluctuating box dipole method (the ratio $\epsr/\epsb$ extracted from fitting the large separation reduction to a constant is given in \Table{tab:tests}).  However, at low $d$, $\Fpolar$ is instead stronger than predicted by classical electrostatic theory: this lack of screening is an artefact of the model, related to the finite size of the solvent molecules.  We note in passing that, to be able to detect this force-reduction effect, low $\epsr/\epsb$ and large $Q/q$ are needed: in the bottom part of \Fig{fig:reduction}, $\epsr/\epsb\simeq 3$, and $Q/q\simeq 63$.

\begin{figure}
  \begin{center}
    \includegraphics[width=3.2in]{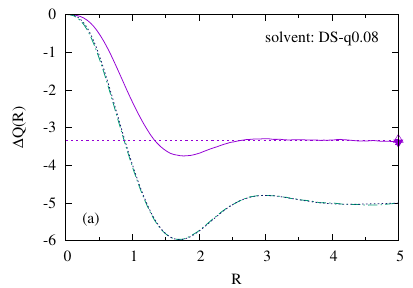}
    \includegraphics[width=3.2in]{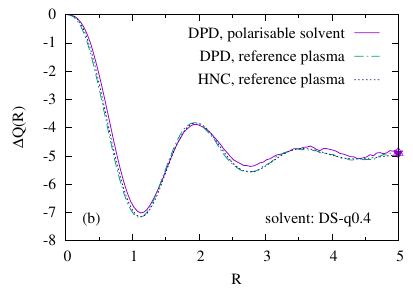}
  \end{center}
  \caption{Solvent charge $\Delta Q(R)$ defined by \Eq{eq:DQR} in the text (solid lines), in a sphere of radius $R$ centered on a test charge $Q=5$, in dressed solvent models with (a) $q=0.08$ and (b) $q=0.4$. Also shown for comparison is $\Delta Q(R)$ for a test charge in a reference plasma, computed by DPD (chain lines) and compared to HNC (dotted lines). As in previous figures, points at large $R$ show the prediction from the fluctuating box dipole method in DPD (open upward triangles) and MC (filled downward triangles).\label{fig:polarization}}
\end{figure}

\subsubsection{Screening around a test charge}\label{subsubsec:testscreen}
The electric field around a test charge $Q$ embedded in our polarisable solvent is partially screened by the dielectric response to $\Evec={Q\rvec}/{4\pi\epsr\epssub0 r^3}$.  Using \Eq{eq:pvec} again, this implies an induced polarisation $\Pvec=(\epsilon-1) Q\rvec/4\pi\epsilon r^3$.  Since $\nabla\cdot(\rvec/r^3) = 4\pi\,\delta^3(\rvec)$, this corresponds to an induced polarisation charge  located exactly on top of the test charge, of magnitude and sign
\begin{equation}
  \Delta Q=-\Bigl(1-\frac{\epsb}{\epsr}\Bigr)\, Q\,.\label{eq:DQ}
\end{equation}
As another test therefore, we compute the solvent charge (\ie\ from the partial charges) contained in a sphere of radius $R$ centered on a positive ion $Q$.  Microscopically, this can be written in terms of the pair distribution functions between the test charge and the partial solvent charges, as
\begin{equation}
  \Delta Q(R) = 4\pi\rhom q \int_0^R\!\!\dr\,r^2
  \,[g_{Q+}(r)-g_{Q-}(r)]\,.\label{eq:DQR}
\end{equation}
We expect this to saturate to the value given in \Eq{eq:DQ} as $R\to\infty$ (\ie\ beyond the dipole-dipole correlation length).

In \Fig{fig:polarization} we show the solvent charge $\Delta Q(R)$ computed as a function of the sphere radius $R$ around a test charge $Q=5$ , for dressed solvent models with small ($q=0.08$) and large ($q=0.4$) permittivities \cite{screennote}. In addition, we also show results for a test charge embedded in the reference plasma of \Sec{subsec:lstheory} (\ie\ the solvent with the partial charges cut loose), and compare this to HNC calculations of the same (\ie\ using a three-component system with a vanishingly small density of test charges).  For all cases it can be seen that after a few damped oscillations $\Delta Q(R)$ becomes constant, attaining the expected value from \Eq{eq:DQ} in the solvent (partial screening) and $-Q$ in the plasma (total screening). The damped oscillations, which persist in the reference plasma and are corroborated by HNC results, reflect the charge ordering taking place in a strong dielectric: see, for example, Keblinski \etal\ \cite{keblinski2000}.  Note that in the negative arms of the oscillations the charge compensation is larger than the test charge itself, $|\Delta Q| > Q$, indicating overscreening locally.

We can use \Eq{eq:DQR} in the limit of large $R$, with \Eq{eq:DQ}, to back out the ratio $\epsilon = \epsr/\epsb$.  This only really works for the case where $\epsilon$ is not too large otherwise we cannot distinguish between partial screening and total screening (\cf\ \partFig{fig:polarization}{b}) which is also the limit of \Eq{eq:DQ} as $\epsr\to\infty$.  Such a situation determines $\epsr/\epsb$ very imprecisely, and we therefore report only the result for $q=0.08$ in \Table{tab:tests}.

\begin{figure}
  \begin{center}
    \includegraphics[width=3.2in]{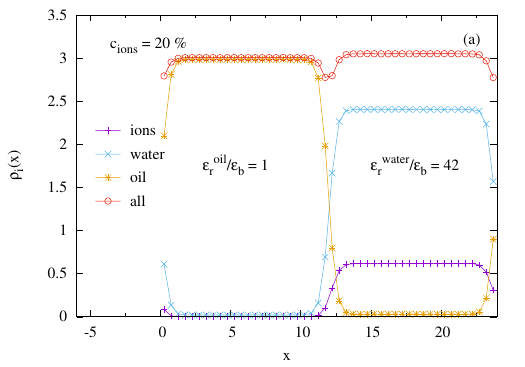}
    \includegraphics[width=3.2in]{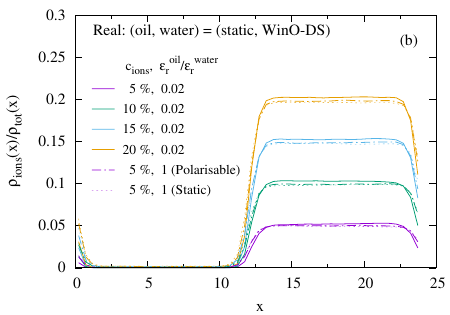}
  \end{center}
  \caption{Desorption of ions from an oil/water interface. Results are from DPD simulations with ions of charge $\pm 1$ and the dressed solvent model with $q=0.36$ or $q=0$. See text for more details.\label{fig:interface}}
\end{figure}

\subsection{Ion desorption from oil/water interface}\label{sec:oilwater}
We now turn to a less trivial example demonstrating that our model shows ion desorption from an oil/water interface of the kind alluded to in the introduction.  We recall from that discussion that a point charge of any sign, embedded in a certain medium, is repelled from an interface with a medium of lower $\epsr$. In the limit of infinite dielectric contrast, it effectively sees an identical image charge across the interface.

Accordingly, as a fourth test, we consider two immiscible solvents, oil and water, and add ions at various concentrations in the water phase. The DPD repulsion parameter is tuned to ensure immiscibility and to prevent the ions from leaving the water phase: $A=40$ between water and oil and between ions and oil (for water and oil, we intend between their neutral beads), whereas for intra-solvent interactions and between ions we keep $A=25$. To reproduce the real dielectric permittivity mismatch between water and oil, we use the dressed solvent model with $q=0.36$ (for water) and with $q=0$ (for oil).

We then observe the ionic density profiles and compare them with controls in which there is no dielectric mismatch, by arranging for the oil and water permittivities to be equal (with both set to the water value).  For these controls, we either set  $q=0$ for both oil and water, with a static background $l_B=1$\,; or set $q=0.36$ for both, with a static background $l_B=42$.

The simulation box is $24\times6\times 6$ and the interface is perpendicular to the $x$-axis, being approximately located at $x=0$. The ions have charges $\pm 1$ to minimize clustering effects. To improve the statistics, for each case we average over 24 runs with different initial spatial configurations.

In \Fig{fig:interface} we show (a) the density profiles for oil (stars), water (crosses) and ions (tilted crosses), for a concentration of 20\% of ions, and (b) a summary of the relative abundance of ions for the four concentrations considered (solid lines), together with the reference results in the polarisable (dot-dashed lines) and static (dotted lines) implementation.  It can be clearly seen that ions are repelled from the oil/water interface when $\epsr^\text{oil}/\epsr^\text{water}<1$ (as compared to $\epsr^\text{oil}/\epsr^\text{water}=1$). Also, it can be seen that the two controls give very similar results.

To quantify this let us suppose Gibbs dividing surfaces are inserted at $x_1$ and $x_2$ such that $0\le x_1< x_2\le L_x$, with the ion-containing water phase occupying the region $x_1\alt x\alt x_2$, and the oil phase occupying the remainder.  Then, noting there are two interfaces in the simulation box, the interfacial excess $\Gamma_i$ (not to be confused with the dimensionless coupling parameter introduced in \Sec{subsec:electrostatics}) for the $i$-th species is given by
\begin{equation}
  2\Gamma_i = \frac{N_i}{A} - \rhoio x_1
  - \rhoiw (x_2-x_1) - \rhoio (L_x-x_2)\,,
\end{equation}
where $N_i$ is the total number of molecules of species $i$ in the simulation box, of cross sectional area $A$, and $\rhoio$ and $\rhoiw$ are the mean densities in the bulk regions of the two phases.  Anticipating that the interfaces are a distance of the order $L_x/2$ apart, and defining $\Delta x = x_2 - x_1 - L_x/2$, this rewrites as
\begin{equation}
  2\Gamma_i = A_i + B_i\,\Delta x\,,
  \label{eq:GibbsGamma}
\end{equation}
where
\begin{equation}
  A_i = \frac{N_i}{A} - \frac{L_x}{2}(\rhoio+\rhoiw)\,,\qquad
  B_i = \rhoio-\rhoiw\,.\label{eq:AB}
\end{equation}
This demonstrates that $\Gamma_i$ depends only on the \emph{separation} of the Gibbs dividing surfaces and not on their absolute positions, and furthermore is a linear function of the relative separation, here expressed as $\Delta x$.  To calculate the coefficients in \Eq{eq:GibbsGamma} we only need to measure the densities of the various species in the bulk regions and insert them into \Eq{eq:AB}.

\begin{figure}
  \begin{center}
    \includegraphics[width=3.2in]{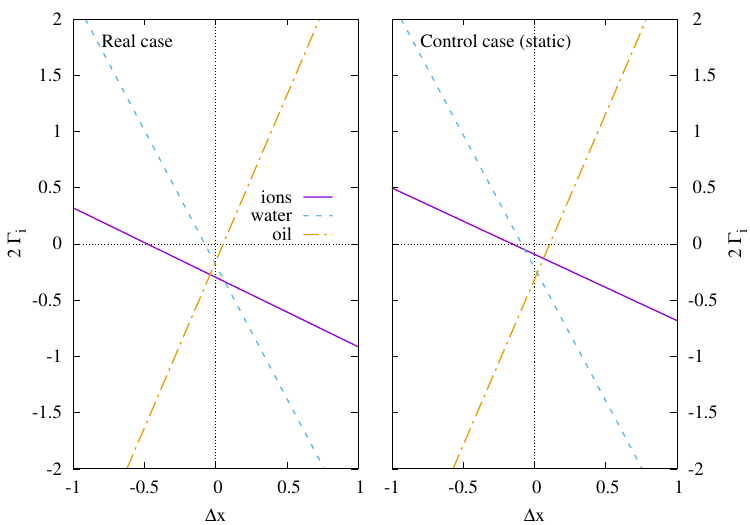}
  \end{center}
  \caption{Interfacial excess for ions, water and oil as a function of the relative separation of the Gibbs dividing surfaces, for a concentration of ions of $20\%$. Left panel: real case, with dielectric mismatch; right panel: control (static) case, with no dielectric mismatch.\label{fig:excess}}
\end{figure}

We do this first for the most concentrated ion solution (20\%), and plot the resulting straight lines in \Fig{fig:excess} as a function of $\Delta x$,  comparing the case where we have a dielectric discontinuity and a fully implemented polarisable water model (on the left) with a `control' case where there is no dielectric discontinuity (on the right).  We first note that there is no unique definition of $\Delta x$ which makes the water and oil interfacial excesses simultaneously vanish.  This corresponds to the `dip' in the total bead concentration seen in \partFig{fig:interface}{a}.  However, most striking is that the line for the ions lies \emph{below} the intersection of the water and oil lines in the left hand plot, but \emph{above} it in the right hand plot.  This is a clear indication that ions are desorped from the interface in the case where there is a dielectric mismatch, compared to the control.

To further quantify this, let us define $\Delta x^*$ as the point where $\Gwater=0$ (\ie\ where the dashed water lines in the plots cross the zero axis).  We can then use this to define the \emph{relative} surface excess for the ions $\Gions^* = \Gions(\Delta x^*)$ \cite{WR}.  This eliminates the dependence on $\Delta x$ and facilitates the propagation of errors in the calculation, however it is still arbitrary to some extent.  Table~\ref{tab:excess} reports the results for a number of cases, clearly confirming ion desorption is significantly present \emph{only} where there is a dielectric mismatch.

\begin{table}
  \begin{ruledtabular}
    \begin{tabular}{lccc}
      $c$ ($\%$) & $\Gions^*$ & control (static) &  control (polarisable) \\[1pt]
      \hline\\[-9pt]
      \phantom{0}5    & $-$0.05(1)\phantom{0}  & $-$0.007(8)  & $-$0.015(4) \\
      10  & $-$0.088(7) & $-$0.013(9) & $-$0.026(8)  \\
      15  & $-$0.107(8) & $-$0.02(1)\phantom{0}  & $-$0.03(1)\phantom{0}  \\
      20  & $-$0.125(8) & $-$0.020(9)  & $-$0.05(1)\phantom{0}  \\[-2pt]
    \end{tabular}
  \end{ruledtabular}
  \caption{Ion surface excesses $\Gions^*$ relative to water (see text) as a function of the concentration in the water phase (total mole fraction of ions, as a percentage).  Note desorption where there is dielectric mismatch (first column), compared to controls (second and third columns).  This is further analysed in \Table{tab:adslen}.\label{tab:excess}}
\end{table}

Finally, for the latter case, we define a (negative) adsorption length by dividing the surface excess by the bulk concentration in the water phase, $\Gions^* /\,\rhobarions$ (we omit the superscript denoting the phase), and compare it to the Debye length $\lD$ calculated from $\lD^{-2}=4\pi l_B \rhobarions$, being sure to use the Bjerrum length for water. Results are given in \Table{tab:adslen} \cite{molarnote}.  We see that the adsorption length is decreasing in magnitude as the concentration increases, and $\Gions^* /\,\rhobarions\approx-\lD/2$.  This latter observation is a little puzzling, since we are in a concentration regime (see, \eg, second column in \Table{tab:adslen}) where $\lD$ should cease to have a physical meaning.  Moreover, the adsorption length is smaller than the charge smearing length, and indeed likely smaller than the width of the interface itself (\partFig{fig:interface}{a}).  Further analysis is beyond the scope of the present study, but this is all highly intriguing and further work to explore this phenomenon is planned.  We can also investigate the effect on the surface tension $\gamma$ itself \cite{gibbsnote}.

\begin{table}
  \begin{ruledtabular}
    \begin{tabular}{cccccc}
      $c$ (\%) & $c/\mathrm{M}$ & $\rhobarions$
      & $\Gions^* /\, \rhobarions$
      & $\lD$
      & $\Gions^* / (\lD\rhobarions)$\\[1pt]
      \hline\\[-9pt]
    \phantom{0}5  & 0.5  & 0.158(2)  &	$-$0.33(9) & 0.71(2) & $-$0.5(1)\phantom{0} \\
    10  & 1.0  & 0.313(1)  & $-$0.28(2) & 0.50(1) & $-$0.56(6) \\
    15  & 1.4  & 0.465(1)  & $-$0.23(2) & 0.41(1) & $-$0.55(6) \\
    20  & 1.9  & 0.617(1)  & $-$0.20(1) & 0.359(9) & $-$0.56(5) \\[-2pt]
    \end{tabular}
  \end{ruledtabular}
  \caption{Adsorption length for ions and the Debye length in the salt solution, and their ratio, for the dielectric mismatch case study (first column in \Table{tab:excess}). The second column indicates the approximate salt concentration in molar units, calculated using the length scale mapping in \Sec{subsec:mapping}.\label{tab:adslen}}
\end{table}

\section{Water-in-vacuum models}\label{sec:waterinvacuum}
We now return to a class of polarisable solvent models which use a vacuum as reference, thus $\epsb=1$. These include the two Peter-Pivkin models\cite{PP14,PLP15}, both of which are based upon the Drude oscillator water model developed for the MARTINI force field \cite{YSS+10} and are equivalent to the dressed solvent model studied in this work with an additional bond angle potential.  As we will discuss in the following, we believe these do not lead to the desired permittivity, therefore also we propose a third corrected version, which is an opportunity to demonstrate our approach to \emph{design} a model with a desired permittivity. The three models are summarised in \Table{tab:drude}, together with a fourth trial one (see below).

\begin{table}
  \begin{center}
    \begin{ruledtabular}
      \begin{tabular}{llllll}
        && Ref.~\onlinecite{PP14}
        & Ref.~\onlinecite{PLP15}
        & *** & \dag\dag\dag\\[1pt]
        \hline\\[-9pt]
        &$r_c$ & $0.646\,\nm$ \\
        &$A$ & 25 \\
        &$\kd/\rc^{-2}$ & $10^5$ \\
        &$d_0/\rc$ & 0.2 & 0.2 & 0.5 & 0.5 \\
        &$k_\theta/\text{rad}^{-2}$ & 1 & 7.5 & 0 & 0 \\
        &$\theta_0$ & 0 \\
        &$q$ & 0.2 & 0.75 & 0.38 & 0.4 \\
        &$\lambda/\rc=1/\betastar$ & 0.7 \\
        &$\lB$ & $55.0\,\nm$ \\
        &$\lB/\rc$ & 85.9 \\
        &$\Gamma$ (coupling) & 1080 \\
        &$\rhom\rc^3$ & 3 \\[3pt]
        \hline\\[-9pt]
        MC & $\myav{\drvec^2}/\rc^2$ & $4.08\times10^{-2}$ &
        $6.41\times10^{-3}$ & 0.412 & 0.406 \\
        & $\gC$ & 1.06(2) & 1.07(3) & 1.20(3) & 1.25(7) \\
        & $\epsr/\epsb$ & 2.87(4) & 5.1(1) & \bf{78(2)} & 89(5) \\
        & $\gK$ & 0.83(2) & 0.78(2) & 0.80(2) & 0.84(4) \\
        \hline\\[-9pt]
        Wert. & $\myav{\drvec^2}/\rc^2$ &
        $4.24\times10^{-2}$ & $8.10\times10^{-3}$ & 0.447 & 0.444 \\
        Ons. & $\epsr/\epsb$ & 3.39 & 7.94 & 104.9 & 115.7 \\[-1pt]
      \end{tabular}
    \end{ruledtabular}
  \end{center}
  \caption{Parameters of Peter-Pivkin models designed for water relative to vacuum, and dielectric properties computed from Monte-Carlo simulations, compared to liquid state theory (\cf\ \Table{tab:dimtri} and \Table{tab:dsprop}). The first two columns show parameters for the two published versions of the model.  The third column (***) shows proposed parameters to recover the dielectric permittivity of water (in bold).  Included for completeness, the fourth column (\dag\dag\dag) is a trial (see text).\label{tab:drude}}
\end{table}

All the models were studied using the same techniques and codes as for the dimer and dressed solvent models, albeit with some modifications. While the dissipative particle dynamics (DPD) code in \DLMESO\ required no modification, a reduction in time step size to $\dt = 0.001$ was needed to explicitly implement the stiff harmonic radial spring forces. The stiffness of these forces required a different approach from the standard Metropolis scheme for the Monte-Carlo (MC) code. In this case, we replace the radial springs by rigid arms of length $d_0$. These arms are free to rotate about the central bead, subject to the additional angular spring (which being soft causes no problems). To make a MC move with these rigid-arm trimers, we first select a bead or partial charge at random. If it is one of the solvent beads, we make a trial linear displacement of the whole molecule.  If it is one of the partial charges we make a trial rotation of the arm, keeping the central bead and other arm fixed in position. The overall scheme is such that detailed balance is satisfied. No significant differences were observed between the DPD calculations with stiff harmonic bonds and MC calculations with rigid arms.

As shown in \Table{tab:drude}, the original Peter-Pivkin models proposed in \Refscite{PP14} and \cite{PLP15} have a dielectric permittivity which is nowhere near as large as reported, and we urge caution in using these models for aqueous systems, recommending alternative values for $d_0$ and $q$ to obtain the dielectric permittivity of water.  Our confidence in these results is bolstered by the generally good alignment we find between the MC and \DLMESO\ simulation results, and the Wertheim and Onsager theories applied to the present situation.

Given that the Bjerrum length is fixed by the choice of background (vacuum), the basic reason why the dielectric permittivity of the original models is too small in the original Peter-Pivkin models is that the dipole moment is too small.  Our proposed remedy is therefore to increase the dipole moment, both by increasing the magnitude of the partial charges and lengthening the distance between them.  If we limit ourselves to the case where the maximum separation between the partial charges $2d_0\le\rc$, this suggests for maximum effectiveness the arm length $d_0=\rc/2$.    We also propose to drop the angular potential as that can only have the effect of reducing the dipole moment.  Then, we only have to adjust the magnitude of the partial charges to get the desired permittivity.

We start by injecting $\epsilon=78$ and $\gK=0.8$ into \Eq{eq:ons}, and utilising the first of \Eqs{eq:ygC} with $\lB/\rc=85.9$ and $\rhom\rc^3=3$, to find we should target $q^2\myav{\drvec^2}\approx 0.0598\,\rc^2$.  We cannot immediately deconvolute this since $q$ enters the plasma correction which affects the Wertheim theory prediction for $\myav{\drvec^2}$.  However, one iteration here will suffice since the error will be corrected in the simulation stage.  Starting from the simple analytic estimate $\myav{\drvec^2}=2d_0^2=0.5\,\rc^2$ from \Appendix{app:intramol}, our initial guess is $q\approx0.345$.  We can now go back and incorporate the plasma correction in the Wertheim theory with this value of $q$.  This improves the estimate to $\myav{\drvec^2}\approx 0.451\,\rc^2$, which now implies (first iteration) $q\approx 0.365$.  Since we know this is unlikely to be exactly correct, let us round this up to $q=0.4$.  If we simulate this (\Table{tab:drude}, final column), we find that $\epsilon\approx89$ which is about 15\% larger than what we want.  Since we expect $\epsilon\propto q^2$ for $\epsilon\gg1$ (\partFig{fig:properties}{a}), we now adjust the magnitude of the partial charges to $q=0.38$ (rounding again).  Checking this revised model, we find that its properties are (perhaps slightly fortuitously) exactly what we want (\Table{tab:drude}, third column).

This approach combines theoretical estimates using Wertheim and Onsager theory, with the assumption $\gK\approx0.8$, to build an initial model which is then simulated to provide a concrete result with a single-step heuristic refinement.  If we did not want to assume $\gK\approx 0.8$ in these models, we could also proceed by first supposing that $\gK=1$, and then measuring the dielectric permittivity and $\gC$ in the initial model to compute a better estimate for $\gK$ from \Eq{eq:fromgC}, that could be fed back into the calculation.  It seems likely that such a procedure should converge after at most one or two iterations.

\section{Discussion}\label{sec:discussion}
In this work we motivated the use of explicitly polarisable solvent models for mesoscale modelling of aqueous structured liquids, and interfaces in aqueous systems.  We have examined two broad classes of such models, with applicability to the well-establised dissipative particle dynamics (DPD) mesoscale simulation methodology.  For water in oil systems ($\epsr/\epsb\approx40$) we propose a specific dressed DPD solvent model `WinO-DS' (\Table{tab:dimtri}) which we confirm behaves as expected in a number of test situations, including capturing ion desorption from an oil-water interface.  Amongst other interesting and worthwhile applications, such as the use of the model to explore ions at solvent interfaces, one can include the self assembly of ionic surfactants \cite{ABDR+18_abbr}, and the effect of a reduced dielectric permittivity in the micelle cores.

Our approach to design these solvent models has been to make sure that the partial charges are only weakly correlated on length scales $\agt\rc$ (representing the solvent granularity), so that above this length scale the solvent appears as a uniform, featureless dielectric continuum.  This motivates the use of molecular dipoles in which the separation of the partial charges is of the order  $\rc$.  Then, the magnitude $q$ of the partial charges is tuned as the final step to set $\epsr/\epsb$.  In all the systems studied, the Kirkwood factor $\gK\approx0.7$--0.8. If we assume that this holds true generally, then a good estimate of the system parameters needed to attain a target relative permittivity could be made via a  combination of Wertheim and Onsager theory.  This can be used as a starting point to iterate towards a final target value of the permittivity, with hopefully only a \emph{single} calibration run required to fine-tune $q$ to get the desired ratio $\epsr/\epsb$.  We therefore anticipate this approach is generically useful, for building mesoscale polarisable solvent models.

The models examined in the present work are all based on molecular dipoles with explicit (smeared) partial point charges.  An intriguing opportunity exists though to make a polarisable solvent model in which true dipoles are embedded into DPD solvent beads.  Most of the technical apparatus for this has already been developed. Ewald summation methods for computing forces and torques on dipoles \cite{S98, AM03, LH08} can presumably be adapted to accommodate charge smearing; and likewise the necessary tools exist to deal with the rotational dynamics of the dipoles, and angular momentum conservation \cite{MFG15}.  If such a solvent model is combined with many-body DPD \cite{War03}, in principle one can simulate vapour-liquid interfaces with dielectric liquids, with possible applications for example to the electrical behaviour of nano-droplets in aerosols.  We leave development along these lines for future work.

\section*{Acknowledgements}
S.C. and M.S. acknowledge support from the European Union's H2020 research and innovation programme under grant number 676531 (Project E-CAM). The STFC/UKRI SCARF supercomputer was used for some of the DPD calculations on the water/oil interface. S.C. thanks A. M. Elena for kindly providing a script to enable better use of resources on SCARF.

\appendix

\section{Intramolecular potentials}\label{app:intramol}
We give details for the intramolecular potentials that can be used to calculate $\myav{\drvec^2}$ in \Sec{subsec:lstheory}, for the models discussed in the main text.  Recall $\beta\phibond_{+-}(r)$ is the intramolecular potential between the partial charges separated by $r$ and arising from the bonded interactions, after integrating out all the other molecular degrees of freedom.

In the dimer class, there are no internal degrees of freedom and the intramolecular potential is trivially $\beta\phibond_{+-}(r)=\half \kd(r-d_0)^2$, where we currently consider the case $d_0 = 0$. Note that unlike many molecular dynamics force fields \cite{pronk2013}, it is conventional in DPD modeling to retain the pairwise soft repulsions between all beads, including bonded pairs.

For the trimer class, one has to integrate out the central bead degrees of freedom.  In general this leads to analytically intractable convolutions of Boltzmann factors, but conveniently for us analytic solutions can be given for the two specific models that we have focussed on in the present work.  The first is the case where there are Hookean springs with $d_0=0$ and no angular potential.  In that case the convolution also gives rise to an effective Hookean spring, with an effective spring constant which is the harmonic mean of the two original spring constants (Hookean springs in series), thus specifically in this case the effective spring constant is $\kd/2$ and the intramolecular potential $\beta\phibond_{+-}(r)= \kd r^2/4$.  For this, neglecting the plasma correction yields $\myav{\drvec^2}=3\kT/(\kd/2)$.

The other case is the Peter-Pivkin model with rigid arms (of length $d_0$) where the only internal degree of freedom of the molecule is the opening angle $\theta$.  Formally
\begin{multline}
  4\pi r^2\exp[-\beta\phibond_{+-}(r)]
  =\int_0^\pi\!\!\dtheta\;\frac{\sin\theta}{2}\,
  \exp[-\smallhalf k_\theta(\theta-\theta_0)^2]\\
    \times \delta(r-2d_0\sin\smallhalf\theta)\,.
\end{multline}
The factors in the integrand are the density of states associated with the angular integration, the Boltzmann weight of the angular potential, and a $\delta$-function constraining the separation between the partial charges to $r$ (see \partFig{fig:models}{b} for geometry).

We make a change of variable to $u=2d_0\sin\smallhalf\theta$. The integral can now be done to get
\begin{equation}
    4\pi r^2\exp[-\beta\phibond_{+-}(r)]
  =\frac{r}{2d_0^2}
  \exp[-\smallhalf k_\theta(\theta-\theta_0)^2]\,,
\end{equation}
where $\theta=2\sin^{-1}(r/2d_0)$.  One application of this is to the case where there is no angular potential and the plasma correction is neglected.  Then $4\pi r^2\prob(r)=r/2d_0^2$ for $r\le2d_0$, and vanishes for $r>2d_0$, so that $\myav{\drvec^2}=\int_0^{2d_0}\!\dr\,r^3/2d_0^2=2d_0^2$.

\section{Charge smearing models}\label{app:ewald}
For completeness we report here details of the charge smearing models used in the simulations, and comment on the link to the Ewald method for dealing with electrostatics in periodic simulation boxes \cite{FS02, IMT09, WVA+13, VJT17}.  We also report in \Table{tab:benchmark} some precision MC results for Slater charges as a benchmark for this problem.  Excellent agreement is found with the HNC liquid state theory, with deviations only becoming apparent in the pressure for the largest $q$ value.

The Ewald method is based on the following summation identity \cite{Kit56, FS02}
\begin{multline}
  \sum_{i>j}\frac{q_iq_j}{r_{ij}}=
    \frac{2\pi}{V}\sum_k A_k |Q_\kvec|^2
    +\sum_{i>j}\frac{q_iq_j}{r_{ij}}\erfc(\alpha r_{ij})\\
      -\frac{\alpha}{\sqrt\pi}\sum_i q_i^2\,,
\end{multline}
where $A_k=e^{-k^2/4\alpha^2}\!/k^2$ and $Q_\kvec=\sum_i q_i\,e^{-i\kvec\cdot\rvec_i}$.  Here $\alpha$ is the Ewald `splitting parameter'.

For charge smeared models this becomes
\begin{multline}
  \sum_{i>j}\frac{q_iq_j}{r_{ij}}[1-f(r_{ij})]=
    \frac{2\pi}{V}\sum_k A_k |Q_\kvec|^2\\
    +\sum_{i>j}\frac{q_iq_j}{r_{ij}}[\erfc(\alpha r_{ij})-f(r_{ij})]
      -\frac{\alpha}{\sqrt\pi}\sum_i q_i^2\,,\label{eq:ewald}
\end{multline}
where $f(r)=\erfc(r/2\sigma)$ for Gaussian charges \cite{WVA+13}, and $f(r)=(1+\betastar\rstar)e^{-2\betastar\rstar}$ for Slater smearing \cite{GMV+06}.  We note the latter is an approximation for the interaction between Slater charges \cite{WV14}, but it is the one that is commonly used in the literature \cite{GMV+06, IMT09, VJT17}. The three terms on the right hand side of \Eq{eq:ewald} are the reciprocal space term, the real space term, and the self energy term. The last is a constant but should be retained to compare with HNC calculations of the energy density.

It is clear that choosing $2\alpha\sigma=1$ for Gaussian charges makes the real space term vanish \cite{CHK11a, WVA+13}. This can be convenient albeit at the expense of losing the ability to tune $\alpha$ for computational efficiency \cite{FS02}.  The final results should be insensitive to the choice of $\alpha$ (\Table{tab:benchmark}).

\begin{table}
  \begin{center}
    \begin{ruledtabular}
      \begin{tabular}{ldddcdd}
        \\[-6pt]
        \multicolumn{7}{l}{$q=0.2\>\to\>\lB q^2=3.436\,\rc$}\\[3pt]
        & \multicolumn{1}{c}{$\alpha$}
        & \multicolumn{1}{c}{$\rcut$}
        & \multicolumn{1}{c}{$\kcut$}
        & & \multicolumn{1}{c}{$\press$}
        & \multicolumn{1}{c}{$\energy$} \\[3pt]
        HNC &      &     &    & & 5.1262 & -13.316\\[3pt]
        MC & 0.75 & 3.0 & 4.5 & & 5.12 & -13.3 \\
        & 1.0  & 3.0 & 5.5 & & 5.13 & -13.3 \\[3pt]
        \hline\\[-6pt]
        \multicolumn{7}{l}{$q=0.52\>\to\>\lB q^2=23.2274\,\rc$}\\[3pt]
        & \multicolumn{1}{c}{$\alpha$}
        & \multicolumn{1}{c}{$\rcut$}
        & \multicolumn{1}{c}{$\kcut$}
        & & \multicolumn{1}{c}{$\press$}
        & \multicolumn{1}{c}{$\energy$} \\[3pt]
        HNC &      &     &     & & 3.0266 & -95.584\\[3pt]
        MC & 0.75 & 3.0 & 4.5 & & 3.03 & -95.6 \\
        & 1.0  & 3.0 & 5.5 & & 3.04 & -95.6 \\[3pt]
        \hline\\[-6pt]
        \multicolumn{7}{l}{$q=0.75\>\to\>\lB q^2=48.3188\,\rc$}\\[3pt]
        & \multicolumn{1}{c}{$\alpha$}
        & \multicolumn{1}{c}{$\rcut$}
        & \multicolumn{1}{c}{$\kcut$}
        & & \multicolumn{1}{c}{$\press$}
        & \multicolumn{1}{c}{$\energy$} \\[3pt]
        HNC &      &     &     & & 1.4927 & -201.372\\[3pt]
        MC & 0.75 & 3.0 & 4.5 & & 1.69 & -201.6 \\
        & 1.0  & 3.0 & 5.5 & & 1.67 & -201.6 \\[-1pt]
      \end{tabular}
    \end{ruledtabular}
  \end{center}
  \caption{Precision Monte-Carlo (MC) results (dimensionless pressure $\press$ and energy density $\energy$) supported by HNC calculations, for a Slater charge plasma under the stated conditions, with $\lB/\rc=85.9$ (coupling parameter $\Gamma=1080$), $A=0$, $\rho\rc^3=6$, $\lambda/\rc=1/\betastar=0.7$.  MC results are reported with an estimated accuracy of 3 significant figures. The HNC pressure is the virial pressure.\label{tab:benchmark}}
\end{table}

The reciprocal and real space sums in \Eq{eq:ewald} need cut-offs, $\kcut$\ and $\rcut$ respectively, and care should be taken with this. For Slater charges with $1/\betastar=0.7$, examination of the real space contribution suggests $\alpha=1$ leads to the smallest $\rcut$, however this is a compromise since a larger $\alpha$ requires a larger $\kcut$ (essentially, this is controlled by the form of $A_k$).

The MC results in \Table{tab:benchmark} were generated for two choices of $\alpha$, with the corresponding rather conservative choices for $\kcut$.  All simulations were done with same real space cut-off $\rcut=3.0$ (in units of $\rc$) which is undoubtedly also rather conservative.  Note that to use the minimum image convention in periodic boundary conditions the box size should be $\ge2\rcut$ (the results in \Table{tab:benchmark} were computed in an $8^3$ box). We note that the pressure is quite sensitive to $\kcut$ as a consequence of an extra factor $k^2$ in the reciprocal space contribution (see Eq.~(12) in \Refcite{WVA+13}).  The real space term contributes to the virial pressure in the usual way \cite{S87}.

Finally, we report the force law that derives from the above Ewald methodology. The general expression for the electrostatic force acting on particle $j$ due to reciprocal and real space contributions is
\begin{equation}
  \begin{split}
  \frac{\myvec{F}_{i}}{\lB q_i} &=
  -\frac{4 \pi}{V} \sum_k i \kvec e^{-i \kvec\cdot\rvec_i} A_k Q_\kvec \\
  &\hspace{2em}{}+ \sum_{j\ne i} \frac{q_j}{r_{ij}^2}
  \Bigl[\frac{2 \alpha r_{ij}}{\sqrt{\pi}} e^{-\alpha^2 r_{ij}^2}
    + \erfc(\alpha r_{ij})\\
  &\hspace{8em}{}+ r_{ij}\,f'\!(r_{ij})-f(r_{ij}) \Bigr]
  \frac{\rvec_{ij}}{r_{ij}}\,,
  \end{split}
\end{equation}
where $f'\!(r)=df\!/dr$, and noting that no contribution arises from the self-energy term. As with the potential, the choice of $\alpha = 1/(2\sigma)$ for Gaussian charges cancels out the real space force contribution. In the case of Slater charge smearing, the expression for the force is correctly given in \Refcite{VJT17}. A previous expression for the force in the literature \cite{IMT09} was based on the derivative of an incorrect form of the second (real space) term in \Eq{eq:ewald}: $\sum_{i>j} ({q_i q_j}/{r_{ij}}) \erfc \left(\alpha r_{ij}\right) \left[1 - f (r_{ij}) \right]$.


\begin{thebibliography}{55}
\bibitem [{sub()}]{subnote}%
  \BibitemOpen
  \href@noop {} {}\bibinfo {note} {With the exception of $\epssub0$, we use
  subscripts to indicate permittivities which are defined relative to free
  space, and no subscript for permittivities defined relative to the static
  background.}\BibitemShut {Stop}%
\bibitem [{oil()}]{oilnote}%
  \BibitemOpen
  \href@noop {} {}\bibinfo {note} {The oil in this case would correspond to a
  non-polarisable liquid with $\epsilon=1$ so that its full permittivity is
  $\epsb\epssub0$.}\BibitemShut {Stop}%
\bibitem [{box()}]{boxnote}%
  \BibitemOpen
  \href@noop {} {}\bibinfo {note} {We use the subscript `box' to distinguish
  this quantity from the polarisation density, introduced in
  \Sec{subsec:tests}.}\BibitemShut {Stop}%
\bibitem [{g11()}]{g110note}%
  \BibitemOpen
  \href@noop {} {}\bibinfo {note} {We have investigated the effect of
  the choice of boundary conditions on $g_{110}(r)$ and plan to publish this
  separately.}\BibitemShut {Stop}%
\bibitem [{scr()}]{screennote}%
  \BibitemOpen
  \href@noop {} {}\bibinfo {note} {We use a pair of equal and
  opposite test charges, held fixed at a distance $5\sqrt{3}\,r_c\approx
  8.66\,r_c$ from each other along the diagonal of a $10\times10\times10$
  (units $\rc$) simulation box.}\BibitemShut {Stop}%
\bibitem [{mol()}]{molarnote}%
  \BibitemOpen
  \href@noop {} {}\bibinfo {note} {To obtain the molar concentrations, we
  divide the total ion density by $2\NA\rc^3$, where the factor `2' arises
  because each neutral ion pair contributes two ions, and $\NA\rc^3$ is taken
  from \Eq{eq:narc3}.}\BibitemShut {Stop}%
\bibitem [{gib()}]{gibbsnote}%
  \BibitemOpen
  \href@noop {} {}\bibinfo {note} {In this context, if $\mu_i$ is the
  chemical potential of the $i$-th species and $\press$ is the pressure, the
  Gibbs-Duhem relation $\sum_i\rho_i\,\dmu_i = \leibnizd\press$ implies $\sum_i
  B_i\,\dmu_i = \sum_i\leibnizd(\press^{(1)}-\press^{(2)})=0$, since by
  mechanical force balance the pressures in the two phases are equal.
  This implies the Gibbs adsorption isotherm $\dgamma=-\sum_i\Gamma_i\,\dmu_i$
  is independent of the choice of position of the Gibbs dividing surfaces, as
  is expected.}\BibitemShut {Stop}%
\end{thebibliography}
\end{document}